\documentclass[prl,twocolumn,superscriptaddress,showpacs,a4paper]{revtex4}
\usepackage{amsmath}
\usepackage{amsfonts}

\begin{document}

\title{Quantum Encodings in Spin Systems and Harmonic Oscillators}
\author{Stephen D.\ Bartlett} 
\email{bartlett@ics.mq.edu.au}
\affiliation{Department of Physics, Macquarie University, Sydney, New
  South Wales 2109, Australia} 
\author{Hubert \surname{de Guise}}
\affiliation{Department of Physics, Macquarie University, Sydney, New
  South Wales 2109, Australia} 
\affiliation{Department of Physics,
  Lakehead University, Thunder Bay, Ontario, P7B 5E1, Canada}
\author{Barry C.\ Sanders} 
\affiliation{Department of Physics,
  Macquarie University, Sydney, New South Wales 2109, Australia}
\affiliation{Erwin Schr\"odinger International Institute for
  Mathematical Physics, Boltzmanngasse 9, A--1090 Vienna, Austria}
\date{April 12, 2002}

\begin{abstract}
  We show that higher--dimensional versions of qubits, or qudits, can be
  encoded into spin systems and into harmonic oscillators, yielding
  important advantages for quantum computation. Whereas qubit--based
  quantum computation is adequate for analyses of quantum vs classical
  computation, in practice qubits are often realized in
  higher--dimensional systems by truncating all but two levels, thereby
  reducing the size of the precious Hilbert space. We develop natural
  qudit gates for universal quantum computation, and exploit the
  entire accessible Hilbert space.  Mathematically, we give
  representations of the generalized Pauli group for qudits in coupled
  spin systems and harmonic oscillators, and include analyses of the
  qubit and the infinite--dimensional limits.
\end{abstract}

\maketitle

Quantum computation may be able to perform certain tasks more
efficiently than a classical computer; for example, Shor's
algorithm~\cite{Sho94} for factoring prime numbers on a quantum
computer is exponentially faster than any known algorithm on a
classical computer.  The standard model of a quantum computer involves
coupling together two--level quantum systems (qubits) such that the
Hilbert space of the system grows exponentially in the number of
qubits.

A major obstacle to universal quantum computing is the limit on the
number of coupled qubits that can be achieved in a physical
system~\cite{Ste98}.  The use of $d$--dimensional, or \emph{qudit},
quantum computing enables a much more compact and efficient
information encoding than for qubit computing.  Qudit quantum
information processing employs fewer coupled quantum systems: a
considerable advantage for the experimental realization of quantum
computing.  The harmonic oscillator is a system that naturally
provides qudits as quanta in its energy spectrum.  Qubits are obtained
by restricting the dynamics to just two of these quanta, namely the
vacuum state $|0\rangle$ and the first excited state $|1\rangle$;
e.g., photons in cavity QED~\cite{Gio00} and
interferometry~\cite{Kni01}.  However, the control of entanglement in
larger Hilbert spaces is now feasible (e.g., orbital angular momentum
states of photons~\cite{Mai01}).  Our aim is to show that the
restriction to two--dimensional Hilbert spaces is not necessary and
that higher--dimensional Hilbert spaces are an advantage, particularly
when the number of achievable coupled systems is limited and
entanglement between systems with larger Hilbert spaces is physically
possible.

A quantum computer also requires gates, realized as the unitary
evolution under some Hamiltonian.  For qubits, a universal set of
gates is given by arbitrary SU(2) rotations of a single qubit along
with some nonlinear coupling transformation between adjacent qubits
generated by a two--qubit Hamiltonian~\cite{Nei00}.  For qudit quantum
computation, the issue of creating a universal set of gates is more
involved.  In particular, it is not possible to treat coupled qudits
as a collection of qubits, because (typically) one does not have
access to ``pairwise'' Hamiltonians between two arbitrary levels of
coupled qudits.  For example, in a system of coupled oscillators
realized as radiation modes in a cavity, Hamiltonians that generate
single--gate operations such as a coupling of the $i^{\text{th}}$
level of one oscillator and the $j^{\text{th}}$ level of another
cannot be realized physically.  Thus, quantum computation with qudits
requires an investigation not only into the coupling of multilevel
systems but also the set of physically realizable Hamiltonians with
which one can construct a universal set of gates.  In this
communication, we develop transformations for a collection of coupled
$d$--level systems.  These transformations are obtained as two
mathematical realizations of a basis of unitary operators for a single
qudit.  We show how each of these realizations can be implemented
either in a spin system or a harmonic oscillator~\footnote{An
  inequivalent encoding of a qudit in a harmonic oscillator is given
  in~\cite{Got01}.}.  We establish a SUM gate~\cite{Got01}, which
couples qudits and serves as the qudit analogue of the controlled NOT
gate; this SUM gate employs a standard two--mode coupling Hamiltonian.

The theoretical investigation of qudit computation is best expressed
in terms of the \emph{generalized Pauli group} for qudits.  Recalling
the Pauli group for a two--level system, a qubit is realized as a
state in a two--dimensional Hilbert space $\mathcal{H}_2$, spanned by
two normalized orthogonal states, $|0\rangle$ and $|1\rangle$, that
serve as a computational basis for $\mathcal{H}_2$.  The unitary
operators $\{ X_2 \equiv \sigma_x,\, Z_2 \equiv \sigma_z \}$, where
$\sigma_i$ is a Pauli spin matrix, generate the Pauli group using
matrix multiplication: the elements of this group are known as Pauli
operators and provide a basis of unitary operators on $\mathcal{H}_2$.

A qudit is realized as a state in a $d$--dimensional Hilbert space
$\mathcal{H}_d$, with a computational basis $\{ |s\rangle; \,
s=0,1,\ldots,d-1 \}$ serving as the generalization of the binary basis
$\{ |0\rangle, \, |1\rangle \}$ of the qubit.  A basis for unitary
operators on $\mathcal{H}_d$ is given by the \emph{generalized Pauli
  operators}~\cite{Pat88,Got01}
\begin{equation}
  \label{eq:PauliOperators}
  (X_d)^a (Z_d)^b , \quad a,b \in 0,1,\ldots d-1 \, ,
\end{equation}
where $X_d$ and $Z_d$ are defined by their action on the computational
basis as follows:
\begin{align}
  \label{eq:ActionOfXOnCompBasis}
  X_d | s \rangle &= | s+1\ (\text{mod}\ d) \rangle \, , \\
  \label{eq:ActionOfZOnCompBasis}
  Z_d | s \rangle &= \exp (2 \pi \text{i} s/d) | s \rangle \, . 
\end{align}
The operators $X_d$ and $Z_d$ generate the noncommutative
generalized Pauli group under matrix multiplication, satisfying
\begin{equation}
  \label{eq:NonCommutivityOfXZ}
  Z_d X_d = \exp(2\pi \text{i}/d) X_d Z_d \, .
\end{equation}

The analysis of the generalized Pauli group as operators in spin
systems and harmonic oscillators is necessary for realizing qudit
algorithms and error correcting codes~\cite{Got01,Got98}.  For spin
systems, we construct the generators of the generalized Pauli group in
$d=2j+1$ dimensions using operators that are expressed in terms of the
SU(2) angular momentum and phase operators.  This construction allows
us to conveniently view a qudit as the Hilbert space of a
$d$--dimensional irreducible representation (irrep) of SU(2).  For
qudits in a harmonic oscillator, we obtain a generalized Pauli group
generated by the number operator $\hat{N}$ and a phase operator
$\hat{\theta}$.  A second realization of qudits is given in terms of
phase states; this realization is ``dual'' to the first realization
given here, and allows for the construction of a simple SUM gate.  By
investigating the $d\to\infty$ limit, we show that it is \emph{not}
the common generalization of the Pauli group for continuous--variable
quantum information (i.e., the Heisenberg--Weyl group) with position
eigenstates as the computational basis.

We begin by constructing a realization of the generalized Pauli group
for a spin system; i.e., in the $d$--dimensional Hilbert space of a
SU(2) irrep of highest weight (angular momentum) $j= (d-1)/2$.
Consider the standard basis for the su(2) algebra $\{ \hat{J}_z,
\hat{J}_\pm = \hat{J}_x \pm \text{i} \hat{J}_y \}$.  Let $\{ |j,m)_z;
\, m=-j,\ldots, j \}$ denote the standard weight basis for the Hilbert
space $\mathcal{H}_{d=2j+1}$ for an SU(2) irrep of highest weight
(angular momentum) $j$.  We use a simplifying notation, allowing $m$
to take all the integer (or half--integer) values modulo $2j+1$, thus
defining $|j,j+1)_z = |j,-j)_z$.

With the computational basis defined to be
\begin{equation}
  \label{eq:CompBasisSU(2)WeightBasis}
  |s\rangle \equiv |j, j-s)_z \, , \quad s=0,1,\ldots, d-1\, ,
\end{equation}
the generators of the generalized Pauli group can be expressed in
terms of operators that act in a natural way on SU(2) basis states.
Because the basis states are eigenstates of $\hat{J}_z$, we have
\begin{align}
  \label{eq:su(2)RepOfX}
  X_d &\mapsto \sum_{m=-j}^j |j,m)_z (j,m+1| \, , \\
  \label{eq:su(2)RepOfZ}
  Z_d &\mapsto \exp \bigl( 2\pi \text{i} (j - \hat{J}_z)/d \bigr) \, ,
\end{align}
which are unitary and satisfy
Eqs.~(\ref{eq:ActionOfXOnCompBasis}-\ref{eq:NonCommutivityOfXZ}).

The operators $X_d$ and $Z_d$ are conjugate to one another:
\begin{equation}
  \label{eq:ConjugationByFourierTransform}
  X_d= U^{-1}\cdot Z_d \cdot U\, ,
\end{equation}
where the unitary transformation $U$ is the Fourier transform in
dimension $d$.  It is convenient to view $X_d$ as the exponent of a
Hermitian operator $\hat{\theta}_z$, defined so that $X_d = \exp (2\pi
\text{i}\, \hat{\theta}_z/d)$, just as $Z_d$ is generated by the
operator $\hat{J}_z$.  The operator $\hat{\theta}_z$ is known as a
phase operator~\cite{Vou90} for a spin system.

The generalized Pauli operators $X_d$ and $Z_d$ can also be realized
as operators that act naturally on the space $\mathbb{H}_d$ of
dimension $d$ spanned by harmonic oscillator states of no more than
$d-1$ bosons.  We define the computational basis to be the set of
harmonic oscillator energy eigenstates
\begin{equation}
  \label{eq:CompBasisOscillatorEigenstates}
  |s\rangle \equiv |n = s \rangle_{\text{HO}} \, , \quad s=0,1,\ldots,d-1
   \, ,
\end{equation}
where $\hat{N}|n\rangle_{\text{HO}} = n|n\rangle_{\text{HO}}$.  Again,
we apply the cyclic notation $|d\rangle = |0\rangle$.  Now defining
the generalized Pauli group as operators on this subspace of the
harmonic oscillator, the generators $X_d$ and $Z_d$ are expressed as
\begin{equation}
  \label{eq:HPrepOfXandZ}
  X_d \mapsto \sum_{s=0}^{d-1} |s+1\rangle \langle s| \, , \quad
  Z_d \mapsto \exp( 2 \pi \text{i} \hat{N} /d) \, ,
\end{equation}
which are unitary on $\mathbb{H}_d$.  Again, we view $X_d$
as the exponent of a Hermitian operator $\hat{\theta}_z$, such that
$X_d = \exp (2\pi \text{i}\, \hat{\theta}_z/d)$; the operator
$\hat{\theta}_z$ is the Pegg--Barnett phase operator~\cite{Peg97},
which is well--defined for finite $d$.  We will call this
representation of the generalized Pauli group the \emph{number
  representation}.

An advantage of this explicit realization of $X_d$ and $Z_d$ as
unitary operators on the harmonic oscillator Hilbert space is that it
enables us to explore the $d \to \infty$ limit in a rigorous way; this
limit yields continuous--variable quantum computation.  The limiting
procedure for phase operators has been thoroughly
investigated~\cite{Peg97,Lyn95}.  In this limit, the computational
basis remains the harmonic oscillator energy eigenstates (now
including all states $s = 0,1,\ldots,\infty$), following
Eq.~(\ref{eq:CompBasisOscillatorEigenstates}).  Note that defining the
phase operator on the infinite--dimensional Hilbert space
$\mathbb{H}_\infty$ of the harmonic oscillator presents
challenges~\cite{Lyn95}.  The number operator $\hat{N}$ and the phase
operator $\hat{\theta}_z$ are conjugate in the same sense that
momentum and position are conjugate, but the limit does \emph{not}
yield the usual continuous--variable Pauli group: the Heisenberg--Weyl
group, with position $\hat{x}$ and momentum $\hat{p}$ operators as
generators.  It is also important that the states of the computational
basis for the limiting case remain harmonic oscillator energy
eigenstates, rather than position (or momentum) eigenstates or
squeezed Gaussians, as are commonly used for continuous--variable
quantum computing.

A second realization of $X_d$ and $Z_d$ in the Hilbert space
$\mathcal{H}_d$ for an irrep of SU(2) can be constructed, with a
computational basis given by SU(2) phase states;  this
representation is ``dual'' to the number representation.  Consider the
relation $\text{i} X_2 = \exp ( \text{i}(\pi/2)X_2 )$ for qubits;
i.e., that
\begin{equation}
  \label{eq:OneIsRotationOfZero}
  |1\rangle = X_2 |0\rangle = (-\text{i})e^{\text{i}(\pi/2)X_2}
   |0\rangle \, .
\end{equation}
The Pauli operator $X_2$ has two interpretations, each of which can be
generalized in a different way.  In the number representation, we
interpret $X_2$ as a cyclic number state raising operator $|1\rangle =
X_2 |0\rangle$ and generalize this operator as a cyclic raising
operator.  However, using the relation~(\ref{eq:OneIsRotationOfZero}),
we can also view $X_2$ as a rotation.  (Using the su(2) representation
$X_2 = 2\hat{J}_x$, this rotation is about the $x$--axis.)  Thus, the
state $|1\rangle$ is obtained (up to a phase) by rotating $|0\rangle$
by an angle $\pi$ about the $x$--axis.  The computational basis states
needed for this type of generalization to qudits are SU(2) phase
states and have been investigated by Vourdas~\cite{Vou90} (although
using rotations generated by $\hat{J}_z$ rather than $\hat{J}_x$).
These states form an orthonormal basis for the SU(2) irrep.

Let $\{ |j,m)_x ;\, m=-j,\ldots,j \}$ be the weight basis for an SU(2)
irrep of angular momentum $j = (d-1)/2$, where $\hat{J}_x$
rather than $\hat{J}_z$ is diagonal; i.e., $\hat{J}_x|j,m)_x
= m|j,m)_x$.  For this representation, we define the computational
basis states to be
\begin{equation}
  \label{eq:PhaseStateCompBasis}
  |s\rangle \equiv \begin{cases}
  \frac{1}{\sqrt{d}} \sum_{m=-j}^j \exp ( 2 \pi \text{i} ms/d )
  |j,m)_x &d\ \text{odd,} \\
  \frac{1}{\sqrt{d}} \sum_{m=-j}^j \exp ( 2 \pi \text{i}
  (m+\tfrac{1}{2}) s/d ) |j,m)_x &d\ \text{even.}
  \end{cases}
\end{equation}
These states form an orthonormal basis for
$\mathcal{H}_d$~\cite{Vou90}.  They are referred to as SU(2)
\emph{phase states} because they are eigenstates of a phase operator
for spin systems.

The generalized Pauli operator $X_d$ on this computational basis is
given by
\begin{equation}
  \label{eq:RotationQuditX}
  X_d \mapsto \begin{cases}
  \exp \bigl( 2 \pi \text{i} \hat{J}_x / d \bigr) &d\ \text{odd,} \\
  \exp ( -\text{i}\pi/d ) \exp \bigl( 2 \pi \text{i}
  \hat{J}_x / d \bigr) &d\ \text{even,} \end{cases}
\end{equation}
satisfying Eq.~(\ref{eq:ActionOfXOnCompBasis}).  Note that $(X_d)^d =
\hat{\openone}$ for both $j$ integral and half--integral.  The
generalized Pauli operator $Z_d$ is given by
\begin{equation}
  \label{eq:RotationQuditZ}
  Z_d \mapsto \sum_{s=0}^{d-1} \exp(2\pi\text{i}s/d) |s\rangle\langle
  s| \, ,
\end{equation}
which is unitary and satisfies Eq.~(\ref{eq:ActionOfZOnCompBasis}).
Note that we can express $Z_d$ as the exponent of a Hermitian operator,
\begin{equation}
  \label{eq:RotationQuditXAsExp}
  Z_d = \exp \bigl(2 \pi \text{i}\, \hat{\theta}_x /d \bigr) \, , \quad 
  \hat{\theta}_x \equiv \sum_{s=0}^{d-1} s |s\rangle\langle s| \, ;
\end{equation}
the operator $\hat{\theta}_x$ is a phase operator for a spin system.

Note that this representation of the generalized Pauli group is
``dual'' to the number representation of
Eqs.~(\ref{eq:su(2)RepOfX})-(\ref{eq:su(2)RepOfZ}) in the same sense that the
position and momentum representations of the harmonic oscillator are
dual.  For the number representation, the computational basis states
are eigenstates of $\hat{J}_z$, and the phase operator
$\hat{\theta}_z$ generates the ``ladder'' transformations.  In the
\emph{phase representation} given here, the computational basis states
are eigenstates of the phase operator $\hat{\theta}_x$, i.e., ``phase
eigenstates'', and it is $\hat{J}_x$ which generates the ladder
transformations via rotations about the $x$--axis.  Both of these
representations can be considered as natural generalizations of the
qubit case, because the standard computational basis $|0\rangle =
|\tfrac{1}{2},\tfrac{1}{2})_z$ and $|1\rangle =
|\tfrac{1}{2},-\tfrac{1}{2})_z$ are both eigenstates of $\hat{J}_z$
and phase eigenstates of $\hat{\theta}_x$.

As with the number representation, this phase representation of the
generalized Pauli group can be expressed in a harmonic oscillator
Hilbert space.  Again considering the finite Hilbert space
$\mathbb{H}_d$, the eigenstates of $\hat{J}_x$ (unlike $\hat{J}_z$ of
the number representation) are replaced with harmonic oscillator
number states $|n\rangle$ with a boson number less than $d$.  The
computational basis, then, consists of finite--$d$ phase eigenstates.
The generalized Pauli operators $X_d$ and $Z_d$ are generated by the
number operator and Pegg--Barnett phase operator, respectively.
Again, the $d\to\infty$ limit yields challenging problems: it is well
known that phase eigenstates do not exist in the infinite--dimensional
Hilbert space $\mathbb{H}_\infty$ of the harmonic
oscillator~\cite{Peg97}.  

Despite the issues involving $d\to \infty$ phase operators, universal
qudit quantum computation is well--defined for finite $d$~\cite{Got01}.
In the following, we discuss these requirements in terms of an
optical realization, where the harmonic oscillators are realized as
modes in a cavity; such a realization has been discussed
in~\cite{Bar01}.  However, this realization is formally equivalent to
any oscillator system.

To perform arbitrary unitary transformations on a single oscillator
\emph{efficiently}, one may employ a combination of linear optics,
squeezing, and a nonlinear process such as photon
detection~\cite{Got01} or a nonlinear optical Kerr
interaction~\cite{Llo99}.  Of particular importance is to realize the
Fourier transform operation on a single qudit, which takes number
eigenstates to phase eigenstates and vice versa.  This operation is
the generalization of the Hadamard transformation for qubits; as it is
a unitary transformation on a single oscillator, it can be performed
efficiently as described above.

For quantum computation, we must also realize a gate that performs a
two--qudit interaction.  A simple controlled two--qudit interaction
gate is the SUM gate~\cite{Got01}
\begin{equation}
  \label{eq:SUMGate}
  {\rm SUM}: \ |s_1\rangle_1\otimes|s_2\rangle_2 \mapsto
  |s_1\rangle_1\otimes|s_1+s_2\ (\text{mod}\ d)\rangle_2 \, .
\end{equation}
Consider two oscillators coupled by the four--wave mixing interaction
Hamiltonian $\chi \hat{N}_1 \hat{N}_2 = \chi \hat{a}_1^\dagger
\hat{a}_1 \hat{a}_2^\dagger \hat{a}_2$.  This Hamiltonian for an
optical system describes a four--wave mixing process in which $\chi$
is proportional to the third--order nonlinear
susceptibility~\cite{Mil83}.  Let oscillator $1$ be in a state $|s_1
\rangle_1$ encoded in the number state basis, and let oscillator $2$
be in a state $|s_2 \rangle_2$ encoded in the phase state basis.  This
interaction Hamiltonian generates the transformation
\begin{equation}
  \label{eq:SUMHamiltonian}
  e^{-{\rm i} \chi \hat{N}_1 \hat{N}_2 t} |s_1 \rangle_1 \otimes |s_2
  \rangle_2 = |s_1 \rangle_1 \otimes |(\tfrac{\chi t}{2\pi}) s_1 + s_2\
  (\text{mod}\ d) \rangle_2 \, .
\end{equation}
Thus, with fixed interaction time $t=2\pi \chi^{-1}$, this Hamiltonian
generates the SUM transformation on two qudits.  (Note that a similar
gate can be defined for spin systems using a
$\hat{J}_{z1}\hat{J}_{z2}$--type Hamiltonian~\cite{San89}.)

Quantum computation with multiple qudits could be performed by
coupling several modes in a single cavity; each mode realizes a single
qudit~\cite{Bar01}.  Modes are coupled via a SUM interaction of the
time described above.  Note that the control qudit for the sum
operation must be encoded in the number state basis, and the target
qudit must be in the phase state basis.  The encodings of each qudit
can be swapped (between number and phase state bases) using the
Fourier transform.

In summary, we have presented realizations of qudit quantum
computation in spin systems and harmonic oscillators in terms of
number and phase operators.  The representations of the generalized
Pauli group, viewed in terms of SU(2) or harmonic oscillator
operators, allows for qudits to be explicitly encoded into such systems.
An advantage of this scheme is that the SUM gate employs a standard
two--mode Hamiltonian to couple two qudits.  From a rigorous
mathematical viewpoint, these realizations give natural extensions of
the qubit--based Pauli group, and allow for the investigation of the
$d \to \infty$ limit and continuous--variable quantum computation.

By employing qudits rather that qubits, the full size of the
accessible Hilbert space can be exploited, with the advantage of
requiring fewer coupled systems for a given quantum information
process.  However, the use of qudits requires a different set of
quantum gates than the usual qubit rotations and two--qubit
interactions that are normally assumed.  The realization of a
universal set of gates using linear optics, squeezing, and a nonlinear
interaction is convenient for certain harmonic oscillator systems but
is not unique; an important challenge is to identify the optimal set
of gates for a particular system.  The analysis presented here
provides the necessary theoretical tools for developing qudit quantum
computation in spin systems and harmonic oscillators as a promising
alternative to qubit quantum computation.

\begin{acknowledgments}
  This project has been supported by an Australian Research Council
  Large Grant and by a Macquarie University Research Grant.  SDB
  acknowledges the support of a Macquarie University Research
  Fellowship.  We acknowledge helpful discussions with S.\ L.\ 
  Braunstein, T.\ Rudolph and B.\ T.\ H.\ Varcoe.
\end{acknowledgments}


\begin{thebibliography}{99}

\bibitem{Sho94} P.\ W.\ Shor, \emph{Proceedings, 35$^{\rm th}$ Annual
  Symposium on Foundations of Computer Science} (IEEE Press, Los
  Alamitos, CA, 1994).

\bibitem{Ste98} A.\ Steane, Rept.\ Prog.\ Phys.\ \textbf{61}, 117 (1998).
  
\bibitem{Gio00} V.\ Giovannetti, D.\ Vitali, P.\ Tombesi and A.\ 
  Ekert, \pra \textbf{62}, 032306 (2000).
  
\bibitem{Kni01} E.\ Knill, R.\ Laflamme, and G.\ J.\ Milburn, Nature
  (London) \textbf{409}, 46 (2001).

\bibitem{Mai01} A.\ Mair, A.\ Vaziri, G.\ Weihs, and A.\ Zeilinger,
  Nature (London) \textbf{412}, 313 (2001).
  
\bibitem{Nei00} M.\ A.\ Nielsen and I.\ L.\ Chuang, \emph{Quantum
  Computation and Quantum Information} (Cambridge University Press,
  Cambridge, 2000).
  
\bibitem{Got01} D.\ Gottesman, A.\ Kitaev and J.\ Preskill, \pra
  \textbf{64}, 012310 (2001).
  
\bibitem{Pat88} J.\ Patera and H.\ Zassenhaus, J.\ Math.\ Phys.\ 
  \textbf{29}, 665 (1988).

\bibitem{Got98} D.\ Gottesman, ``Fault--Tolerant Quantum Computation
  with Higher--Dimensional Systems,'' \texttt{quant-ph/9802007}.

\bibitem{Vou90} A.\ Vourdas, \pra \textbf{41}, 1653 (1990).
  
\bibitem{Peg97} D.~T.\ Pegg and S.~M.\ Barnett, J.\ Mod.\ Optics
  \textbf{44}, 225 (1997).
  
\bibitem{Lyn95} R.\ Lynch, Phys.\ Rep.\ \textbf{256}, 368 (1995).
  
\bibitem{Bar01} S.\ D.\ Bartlett, B.\ C.\ Sanders, B.\ T.\ H.\ Varcoe
  and H.\ de Guise, in \emph{Experimental Implementation of Quantum
  Computation (IQC'01)} ed.\ R.\ Clark, (Rinton, Princeton, NJ, 2001),
  pp. 344--347.

\bibitem{Llo99} S.\ Lloyd and S.~L.\ Braunstein, \prl \textbf{82},
  1784 (1999).

\bibitem{Mil83} G.\ J.\ Milburn and D.\ F.\ Walls, \pra \textbf{28},
  2065 (1983).

\bibitem{San89} B.\ C.\ Sanders, \pra \textbf{40}, 2417 (1989).
  
\end{thebibliography}
\end{document}